\documentclass[preprint,12pt]{elsarticle}

\usepackage{epstopdf}
\usepackage{amssymb}
\usepackage{amsmath}
\usepackage{graphicx}
\usepackage{natbib}
\usepackage[version=3]{mhchem}
\usepackage[colorinlistoftodos]{todonotes}
\usepackage{lineno}

\newcommand{\wn}{cm$^{-1}$}

\newcommand{\ch}{CH$^+$}

\begin{document}


\title{High-resolution infrared action spectroscopy of the fundamental vibrational band of CN$^+$} 
%

\author[1]{Jos\'e L. Dom\'enech}%

\author[2]{Oskar Asvany}
\author[2,3]{Charles R. Markus}
\author[2]{Stephan Schlemmer}

\author[2]{Sven Thorwirth\corref{cor1}}
\ead{sthorwirth@ph1.uni-koeln.de}
\cortext[cor1]{Corresponding author}
\address[1]{Instituto de Estructura de la Materia (IEM-CSIC), Serrano 123, E28006 Madrid, Spain}
\address[2]{I. Physikalisches Institut, Universit\"at zu K\"oln, Z\"ulpicher Str.~77, D50937 K{\"o}ln, Germany}
\address[3]{Department of Chemistry, University of Illinois,  Urbana, IL 61801, USA
}

\begin{abstract}
Rotational-vibrational transitions of the fundamental vibrational modes of the $^{12}$C$^{14}$N$^+$ and  $^{12}$C$^{15}$N$^+$ cations 
have been observed for the first time using a cryogenic ion trap apparatus with an action spectroscopy scheme.
The lines $P(3)$ to $R(3)$  of $^{12}$C$^{14}$N$^+$ and $R(1)$ to $R(3)$ of $^{12}$C$^{15}$N$^+$ have been measured, limited by the trap temperature of   approximately 4~K and the restricted tuning range of the infrared laser. 
Spectroscopic parameters are presented for both isotopologues, 
with band origins at 2000.7587(1) and 1970.321(1)~\wn, respectively,
as well as an isotope independent fit combining the new and the literature data.
\end{abstract}

\begin{keyword}
molecular ions \sep high resolution \sep trapped ions \sep interstellar molecules
\end{keyword}

\maketitle

\section{Introduction}
\label{sec:intro}

The first molecules observed in the interstellar medium (ISM) were the carbon containing diatomic species CH, \ce{CH+}, and CN, all identified through their electronic spectra \cite{mckellar_PASP_52_187_1940,adams_ApJ_93_11_1941}.
The cyano radical (CN) has been the subject of many laboratory and astronomical investigations, having been found in a variety of astronomical environments including diffuse and dense molecular clouds \cite{Liszt2001,Boger2005}, circumstellar envelopes \cite{Wootten1982}, and external galaxies \cite{henkel_AA_201_L23_1988}.
For decades, it has been a valuable probe of the conditions in these environments and has been used to accurately determine the temperature of the cosmic microwave background \cite{Field1966,Meyer1985,Liszt2001}.
Its cation form, \ce{CN+},  is thought to play an important role in the formation of CN in shielded regions where there is sufficient \ce{N2} density \cite{VanDishoeck1986,Boger2005}.
In regions with high \ce{H2} density, on the other side, the presence of \ce{CN+} is unlikely, 
due to its fast reaction with molecular hydrogen~~\cite{rak84,hun77,pet91,sco97}.
A preliminary search for \ce{CN+} in the ISM, which relied on predictions of the rotational transitions from ultraviolet measurements~\cite{douglas_ApJ_119_303_1954,lutz_ApJ_163_131_1971}, ultimately proved unsuccessful~\cite{hollis_ApJ_219_74_1978}.

\ce{CN+} is isoelectronic with \ce{C2}, and therefore was expected to have either a $^1\Sigma^+$ or $^3\Pi$ ground state. 
Quantum-chemical calculations have had difficulty predicting the relative energy of these low lying states, and have not been able to definitively determine which electronic state has the lowest energy \cite{Bruna1980,peterson_JCP_102_262_1995}. 
The first laboratory measurements of \ce{CN+} were achieved through ultraviolet emission spectroscopy of \ce{C2N2}/He discharges \cite{douglas_ApJ_119_303_1954,lutz_ApJ_163_131_1971}, 
however, at that time it was not possible to identify the electronic ground state with certainty.
Recently, millimeter- and sub-millimeter wave rotational spectra of both the \ce{C^14N+} and \ce{C^15N+} isotopologues were observed for the first time using 
action spectroscopy via
state-dependent attachment of He-atoms in a cryogenic ion trap~\cite{tho19}, 
finally confirming a $^1\Sigma^+$ ground electronic state.

In the present investigation, we have extended the sub-mm work on \ce{CN+} to the infrared.  
Rotational-vibrational transitions of \ce{C^14N+} and \ce{C^15N+} were predicted from molecular constants derived from a global isotope invariant fit of the sub-mm and UV data \cite[see Ref.][]{tho19}.
A survey of the fundamental vibrational band in the 5~$\mu$m region was conducted using a similar action spectroscopic technique
which employed the method of Laser Induced Inhibition of Complex Growth (LIICG)\cite{asv14}.
Due to its cryogenic operation principle and the small rotational partition function of CN$^+$ at low temperatures, only seven \ce{C^14N+} and three \ce{C^15N+} transitions were observed, 
reaching $J^{\prime\prime}_{max}=3$. 
These are the first infrared measurements of \ce{CN+}.
This investigation provides experimental benchmarks for future high-level computational investigations of \ce{CN+} as well as accurate rest frequencies for future infrared studies and astronomical searches.

\section{Experimental setup}

\begin{figure}
 \includegraphics[width=\columnwidth]{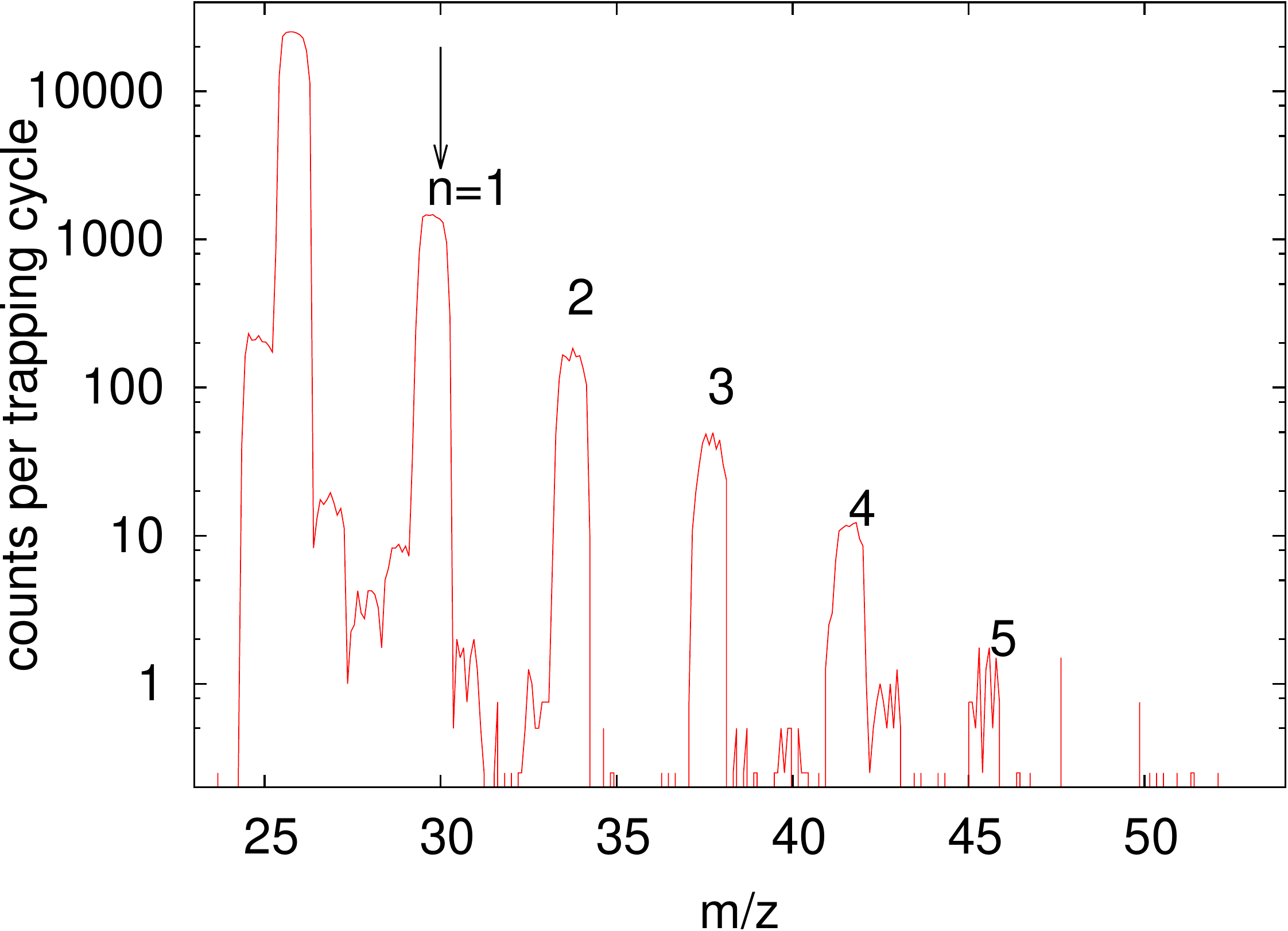}
 \caption{\label{fig1} CN$^+$-(He)$_{\rm n}$ species obtained 
 by trapping $^{12}$C$^{14}$N$^+$ for 1.7~s in a 4~K helium environment.
The negligible  counts for $n=5$  
might be a hint for a shell closure at $n=4$, 
as observed for \ch\ \citep{dom18}. 
The arrow indicates the mass channel ($n=1$) in which the LIICG-detection occurs.
}
\end{figure}

The experiments were conducted in a cryogenic 22-pole ion trap apparatus (COLTRAP), 
which has previously been described in detail ~\citep{asv10,asv14}.
As the recent rotational investigation of CN$^+$  has been performed in the same machine 
with very similar experimental conditions  \cite{tho19}, only a brief description is given here.
CN$^+$ ions are created inside a storage ion source through electron impact ionization ($E_{e^-}\approx 30$~eV) of  methyl cyanide vapour, CH$_3$CN.      
As isotopically enriched methyl cyanide is readily available commercially,
C$^{15}$N$^+$ was produced from CH$_3$C$^{15}$N (Sigma Aldrich, 98~\% $^{15}$N).
Using an excess of helium in the source chamber ($p \approx 10^{-4}$~mbar) turned out to further enhance CN$^+$ production, as described in \cite{tho19}.
Ion pulses are extracted from the source and mass selected
(e.g.\ $m/z=26$ for $^{12}$C$^{14}$N$^+$) in a subsequent quadrupole mass filter. 
When the selected ions enter the 22-pole trap, they are slowed down and cooled to the ambient trap temperature of 4~K by collisions with
He gas (the trap is constantly filled with a number density of about $10^{15}$~cm$^{-3}$). 
Typically,  about 10,000 ions are stored inside the trap, and due to the low temperature and large helium number density, also 
CN$^+$-He complexes are formed via ternary collisions. Because of destruction of these complexes in collisions with He atoms, 
an equilibrium 
\begin{equation}
\label{eq1} 
\rm{CN^+} + 2\,He   \,\,\,  \mathop{\rightleftarrows}    
 \,\,\,  CN^+\text{-}He + He  \,\, , 
\end{equation}
is reached inside the trap after a couple of 10~ms. 
A typical distribution of the CN$^+$-(He)$_n$ complexes formed in the trap 
is shown in Fig.~\ref{fig1}.

For the detection of the rovibrational transitions, the  method  of laser induced inhibition of complex growth 
 (LIICG, \cite{asv14,asv15,sav15,jus17,koh18,dom18,mar19}) was applied.
This action spectroscopic method exploits the fact that  
vibrational excitation of the bare ion leads to 
a suppression of  the formation of the cation-helium complex  (CN$^+$-He in this case).
This action  can be  detected by extraction of the trap content,
mass selection for CN$^+$-He (e.g.,\ at $m/z=26+4=30$, see arrow in Fig.~\ref{fig1}), and counting of these clusters using a Daly-type detector.
A rovibrational line, appearing as dip in the  CN$^+$-He counts, is recorded by repeating these trap cycles at 1~Hz and 
counting CN$^+$-He as a function of the laser frequency.

In the present investigation,
a quantum cascade laser (QCL, Daylight Solutions) operating in the range 1961--2205~cm$^{-1}$
and providing up to several 100~mW of power, has been applied.  
The intrinsic linewidth of the QCL is specified to be smaller than  30~MHz. 
A fraction of the mid-IR light was picked off and guided to a wavemeter (Bristol 621 A-IR) for frequency determination, while the rest was sent through the UHV environment of the 22-pole ion trap via a pair of CaF$_2$ windows. 
After leaving the ion trap machine the radiation passed through a 20~cm long cell containing OCS gas at $\sim1$ mbar pressure. A power sensor head (Thorlabs S302C) was placed at the end of the cell to monitor the power of the IR radiation and to record the OCS absorption lines which were used for absolute frequency calibration. 

   
\section{Rotational-vibrational transitions of \ce{C^{14}N+} and C$^{15}$N$^+$}

\begin{figure}
 \includegraphics[width=\columnwidth]{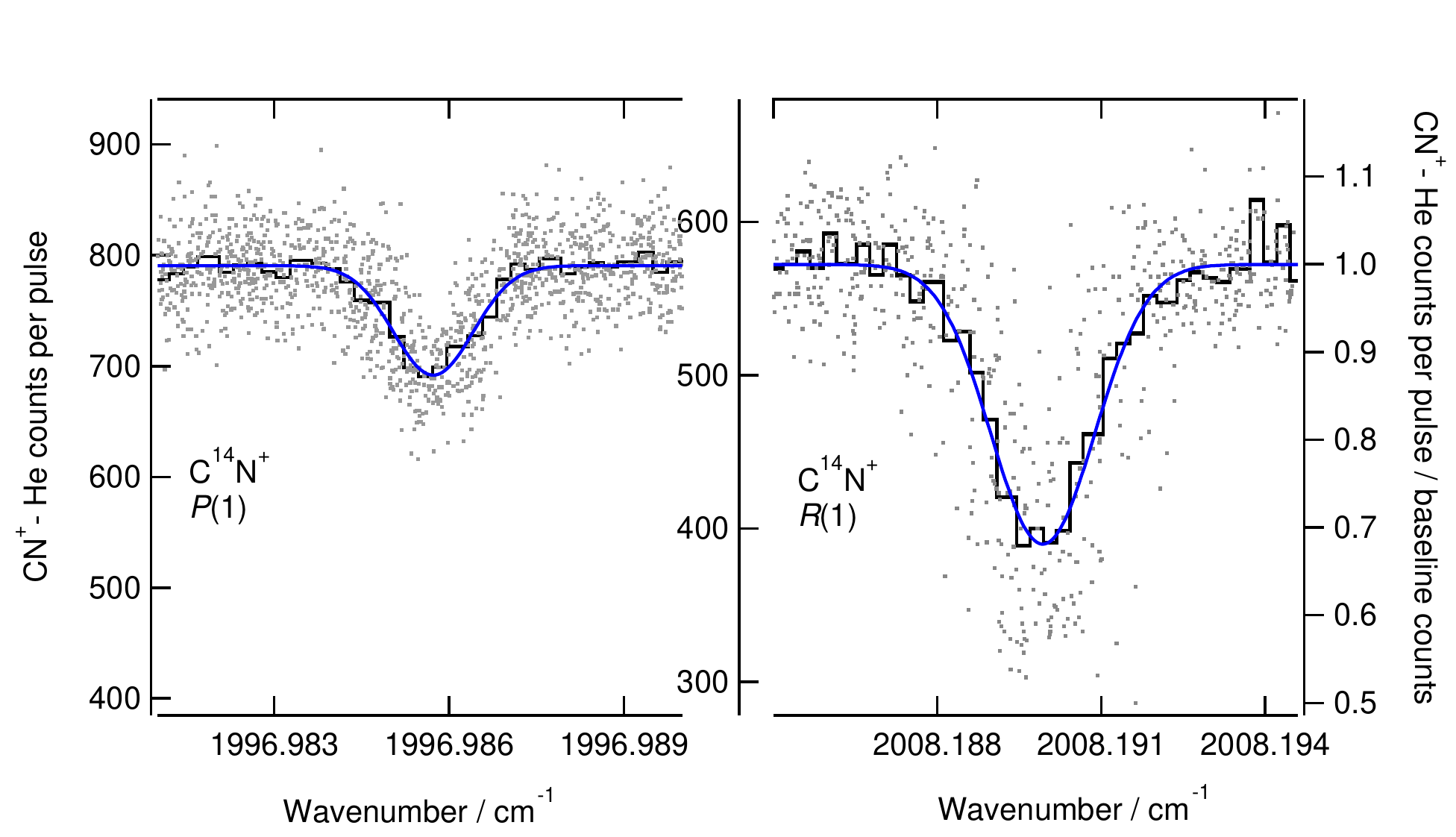}
 \caption{\label{fig2}
 The $P(1)$ and $R(1)$ transitions of the fundamental band of C$^{14}$N$^+$, measured as a depletion of the CN$^+$-He counts, after a trapping time of 740~ms.
 The linewidth is due to Doppler broadening, power broadening, and the linewidth of the QCL.
 The grey points are the ion count data from each trapping cycle 
 (left and center y-axis: absolute counts for the  $P(1)$ and $R(1)$ measurements respectively, right y-axis: normalized counts) 
 and the blue trace is the result from a Gaussian fit.
  From the derived full width at half maximum of 48~MHz for $P(1)$, 
  the effective temperature is 32~K.
 To guide the eye, the average of the data within 8~MHz bins is shown in black. 
}
\end{figure}

\begin{table*}[htbp]  
 \caption{\label{tab1}Wavenumbers of rovibrational transitions of  \ce{C^{14}N+}  and C$^{15}$N$^+$ and fit residuals $o-c$ (in cm$^{-1}$; residuals taken from isotope invariant fit). 
The accuracy of our measurements is constrained by the wavemeter and the upper limit for the quoted accuracy of the calibration lines.  It is expressed in parentheses in units of the last digit.}
 \begin{center}
 \begin{tabular}{rccr@{}lcr@{}l} \hline 
         &       & \multicolumn{3}{c}{\ce{C^{14}N+}} &   \multicolumn{3}{c}{C$^{15}$N$^+$} \\ \cline{3-5} \cline{6-8}
  & $(v',J')  \leftarrow  (v'',J'')$ & exp.  & \multicolumn{2}{c}{$10^4(o-c)$}          &  exp.  & \multicolumn{2}{c}{$10^4(o-c)$}  \\
\hline
$P(3)$   & $(1,2)  \leftarrow  (0,3)$   &  1989.32514(16)  & $-$3. & 0 & --             & \multicolumn{2}{c}{--} \\
$P(2)$   & $(1,1)  \leftarrow  (0,2)$   &  1993.17469(16)  &    2. & 0 & --             & \multicolumn{2}{c}{--} \\
$P(1)$   & $(1,0)  \leftarrow  (0,1)$   &  1996.98573(16)  &    1. & 1 & --             & \multicolumn{2}{c}{--} \\
$R(0)$   & $(1,1)  \leftarrow  (0,0)$   &  2004.49333(16)  & $-$1. & 4 & --             & \multicolumn{2}{c}{--} \\
$R(1)$   & $(1,2)  \leftarrow  (0,1)$   &  2008.18993(16)  &    0. & 7 & 1977.52511(17) &         0. & 5  \\
$R(2)$   & $(1,3)  \leftarrow  (0,2)$   &  2011.84773(17)  &    0. & 4 & 1981.07275(17) &      $-$0. & 5  \\
$R(3)$   & $(1,4)  \leftarrow  (0,3)$   &  2015.46685(19)  &    0. & 5 & 1984.58361(17) &      $-$0. & 0  \\
\hline
 \end{tabular}
 \end{center}
\end{table*}

Seven rovibrational lines of the fundamental vibrational band, from $P(3)$ to $R(3)$, were recorded  for C$^{14}$N$^+$,
and three  rovibrational lines, from $R(1)$ to $R(3)$, for C$^{15}$N$^+$, in total spanning the range from 1977.5 to 2015.5~\wn.  
Limitations in tunability and coverage at the lower wavenumber end of the QCL prohibited
measurements of additional transitions of the C$^{15}$N$^+$ species.
Experimental transition wavenumbers of each  CN$^+$ line were derived from Gaussian fits to the spectral signals, 
as shown in Fig.~\ref{fig2} for the $P(1)$ and $R(1)$ lines.
A minimum of six sweeps across each line was performed in both scanning directions.  
The wavenumber accuracy obtained in this fashion alone is on the order of a few $10^{-3}$\,cm$^{-1}$ 
and is limited by the accuracy of the wavemeter. 
The accuracy was further improved by including a carbonyl sulfide, OCS, reference gas cell.
The scan ranges about the \ce{CN+} transitions included at least one rovibrational OCS transition for 
which the wavenumber is known to very high accuracy as tabulated in the HITRAN database \cite{gordon_JQRST_203_3_2017}. 
The final CN$^+$ line wavenumbers have been calculated as ${\rm \tilde{\nu}(CN^+)=\tilde{\nu}(CN^+_{meas})+(\tilde{\nu}(OCS_{cal}) - \tilde{\nu}(OCS_{meas}))}$.
Therefore, the uncertainty of the final wavenumbers has been estimated as the quadratic mean of i)~the statistical uncertainty from the CN$^+$ line center derived from the fit (ranging from $2-9\times 10^{-5}$ \wn ~depending on the signal-to-noise ratio),  ii)~the rms of the scattered line centers derived from repeated measurements over a CN$^+$ line ($9\times 10^{-5}$ \wn, which reflects the actual repeatability of the wavemeter readings), iii)~the statistical uncertainty of the OCS line centers derived from their fit (ranging from $0.8-1.2\times 10^{-5}$ \wn), iv)~the rms of the scattered line centers derived from repeated measurements over a OCS calibration line ($1\times 10^{-4}$ \wn), and, finally, v)~the uncertainty for the tabulated OCS wavenumbers.  For the latter, we have adopted the wavenumber values and the upper uncertainty limits quoted in the HITRAN2016 database, i.e.\ $9\times 10^{-5}$ \wn\  \citep{gordon_JQRST_203_3_2017}. 
For most lines the accuracy is limited by the wavemeter repeatability 
and by the conservative estimate of the accuracy of the calibration lines.

The transition wavenumbers and their uncertainties are shown in Table~\ref{tab1}
for both isotopologues, while Table~\ref{param1} shows the rotational constants 
and band origins for the $v=1$ levels. These were obtained by fitting the frequency values 
to a linear rotor energy expression using the program PGOPHER \citep{wes17}.
In the fit, the ground state constants ($v=0$)  were kept fixed at the values obtained from
millimeter-/sub-millimeter-wave spectroscopy published previously \cite{tho19}.  
For C$^{14}$N$^+$, the obtained residual of the fit, $\sigma = 4.3$~MHz, is consistent with 
the uncertainties estimated above. For C$^{15}$N$^+$, due to the limited number of measured lines, the centrifugal distortion constant $D_1$ has been fixed to a scaled value.





\begin{table}
\begin{center}
\caption{\label{param1} The best fit spectroscopic parameters of \ce{C^{14}N+}  and C$^{15}$N$^+$ obtained by fitting the data given in Table~\ref{tab1} with the PGOPHER program \cite{wes17}. 
The ground state parameters were fixed to the highly accurate values given in \cite{tho19}.
The numbers in parentheses give the uncertainty of the last digits. See text for more details.}
\begin{tabular}{lr@{}lr@{}ll}
\hline
C$^{14}$N$^+$ 	  & \multicolumn{2}{c}{$v=0$} &     \multicolumn{2}{c}{$v=1$}           & unit \\
\hline
  $\nu$  & \multicolumn{2}{c}{$\cdots$}               &         2000&.7587(1)   &   cm$^{-1}$  \\
  $B_v$    &  56556&.90031(57)     &     55982&.7(12)                     &  MHz \\
  $D_v$    &      0&.199511(46)    &         0&.15(6)                    &  MHz  \\
\hline
\hline
C$^{15}$N$^+$ 	         & \multicolumn{2}{c}{$v=0$} &  \multicolumn{2}{c}{$v=1$}           & unit \\
\hline
  $\nu$ &  \multicolumn{2}{c}{$\cdots$}              &          1970&.321(1)   &   cm$^{-1}$  \\
  $B_v$   &       54826&.28843(49)   &     54272&.5(27)                    &  MHz \\
  $D_v$  	&      0&.187327(32)        &         0&.14 $^a$                 &  MHz  \\
\hline
\end{tabular}
\end{center}
$^a$ fixed to a scaled value
\end{table}


\section{Isotope invariant fitting}

The newly obtained data have also been reduced using a global Dunham-Watson type
model \cite{watson_JMS_80_411_1980} and a revised set of isotope invariant parameters has been obtained complementing
the data set derived earlier \cite{tho19}. A detailed account on the strategies employed in such kind of treatment has been given elsewhere
\cite{bizzocchi_mp_113_801_2014,muller_JPCA_117_13843_2013}.
Fitting was performed using Pickett's SPFIT/SPCAT
suite of programs \cite{pickett_JMolSpectrosc_148_371_1991} and
a truncated output from SPFIT is given as electronic supplementary material to this article.
The new set of isotope independent parameters is given in Table \ref{dunham}, where it is compared
against the parameter set of Ref. \cite{tho19}.  
Inclusion of the new high-resolution infrared data afforded consideration of
the $Y_{21}$ parameter and the term $\delta_{10}^{\rm N}$ which accounts for the breakdown
of the Born-Oppenheimer (BO) approximation. 
This term is defined as $\delta_{10}^{\rm N}=U_{10}\mu^{-1/2}\frac{m_e}{\rm M_N}\Delta^{\rm N}_{10}$, with $U_{10}$ being an isotope invariant term, $\frac{m_e}{\rm M_N}$ the electron-to-nitrogen atomic mass ratio, $\mu$ the reduced mass of \ce{^12C^14N+} and $\Delta^{\rm N}_{10}$ a BO breakdown (BOB) parameter
accounting for nitrogen substitution.
The IR rms of the fit is about 0.1\,cm$^{-1}$, which is a consequence of the limited quality
of the combination differences from the electronic spectra \cite{douglas_ApJ_119_303_1954} used in
the fit. The rms of the new high resolution data in Table~\ref{tab1} alone
is 1.3$\times 10^{-4}$\,cm$^{-1}$. 

From the $Y_{01}(\approx B_e)$ parameter determined here, an experimental 
equilibrium bond length of $r_e=1.1730494(8)$\,\AA\ is derived. This value agrees to
much better than
$10^{-3}$\,\AA\ with the value derived from the initial UV experiments
and to within some $3-4\times 10^{-3}$\,\AA\ with the highest level quantum-chemical calculations
reported in the literature so far \cite[see Ref.][and references therein]{polak_spectrochimacta_58_2029_2002}.


\begin{table*}
\begin{center}
\caption{Isotope invariant fits of CN$^+$.$^a$ \label{dunham}}
\begin{tabular}{lr@{}lr@{}l}
\hline
Parameter  & \multicolumn{2}{c}{Ref. \cite{tho19}} & \multicolumn{2}{c}{This study}      \\ \hline  
$Y_{01}$/MHz                      &   56836. & 013(27)           &  56832. & 336(77)     \\
$Y_{02}$/MHz                      &    $-$0. & 199421(27)        &   $-$0. & 199420(27)  \\
$Y_{11}$/MHz                      &  $-$558. & 227(54)           & $-$543. & 408(284)    \\
$Y_{10}$/cm$^{-1}$                &    2033. & 044(55)           &   2033. & 095(32)$^b$ \\
$U_{10}\mu^{-1/2}$/cm$^{-1}$      & \multicolumn{2}{c}{$\cdots$} &   2032. & 984(31)     \\
$\delta^{\rm N}_{10}$/cm$^{-1}$   & \multicolumn{2}{c}{$\cdots$} &      0. & 1115(95)    \\
$Y_{20}$/cm$^{-1}$                &   $-$16. & 194(22)           &  $-$16. & 168(20)     \\
$Y_{21}$/MHz                      & \multicolumn{2}{c}{$\cdots$} &   $-$3. & 733(68)     \\
$eQq_{00}(^{14}$N)/MHz            &    +4. & 9660(32)            &   +4. & 9660(33)      \\
$C_{00}(^{14}$N)$\times 10^3$/MHz &     4. & 11(65)              &    4. & 13(65)        \\
$rms$(mmw)/MHz                    &     0. & 007                 &    0. & 007           \\
$rms$(IR)/cm$^{-1}$               &     0. & 114                 &    0. & 104           \\
Weighted $rms$                            &     1. & 06                  &    1. & 00            \\ 
\hline
\end{tabular}
\end{center}
$^a$ $Y_{ij}\simeq U_{ij}\times \mu^{-(i+2j)/2}$, see \cite[Ref.][]{watson_JMS_80_411_1980}. \\
$^b$ Derived value.
\end{table*}


 \section{Astrochemical considerations and future perspectives}

Now that the high resolution pure rotational and rotational-vibrational spectra of CN$^+$ have been
studied, a closer look at the reaction kinetics may be indicated. 
Of particular importance in an astrophysical scenario is the reaction of CN$^+$ with molecular hydrogen.
Hydrogenation proceeds in two steps finally leading to the terminal product HCNH$^+$, 
\begin{eqnarray}
\label{eq2} 
\rm{CN^+} +  \rm{H}_2 &    \stackrel{k_2}{\longrightarrow}   &  \rm{HCN^+}   +  \rm{H}  \\
\rm{HCN^+} + \rm{H}_2 &    \stackrel{k_3}{\longrightarrow}   &  \rm{HCNH}^+  +  \rm{H}  \,\, , 
\end{eqnarray}
with a similar reaction chain involving HNC$^+$ as the intermediate
(HCN$^+$ and HNC$^+$ are formed in equal amounts in reaction (2)~\cite{pet91}). 
In the past, these reactions have been investigated at room temperature using SIFT (selected ion flow tube) and ICR (ion-cyclotron resonance) techniques ~\cite{rak84,hun77,pet91,sco97},
and rate coefficients close to the Langevin collision rate were determined
($k\approx 1.54 \times 10^{-9}$~cm$^{3}$s$^{-1}$ for both reaction steps). 
Consequently, this behaviour 
will make detection of CN$^+$ in denser regions of space challenging. 
To the best of our knowledge, there are no measurements of these reactions 
at the cryogenic temperatures prevailing in the ISM. 
For such exothermic and direct hydrogen abstraction reactions,
no temperature dependence is expected.
Nonetheless, we caught a preliminary glimpse of this
reaction chain at a nominal trap temperature of 15~K.
\begin{figure}
 \includegraphics[width=\columnwidth]{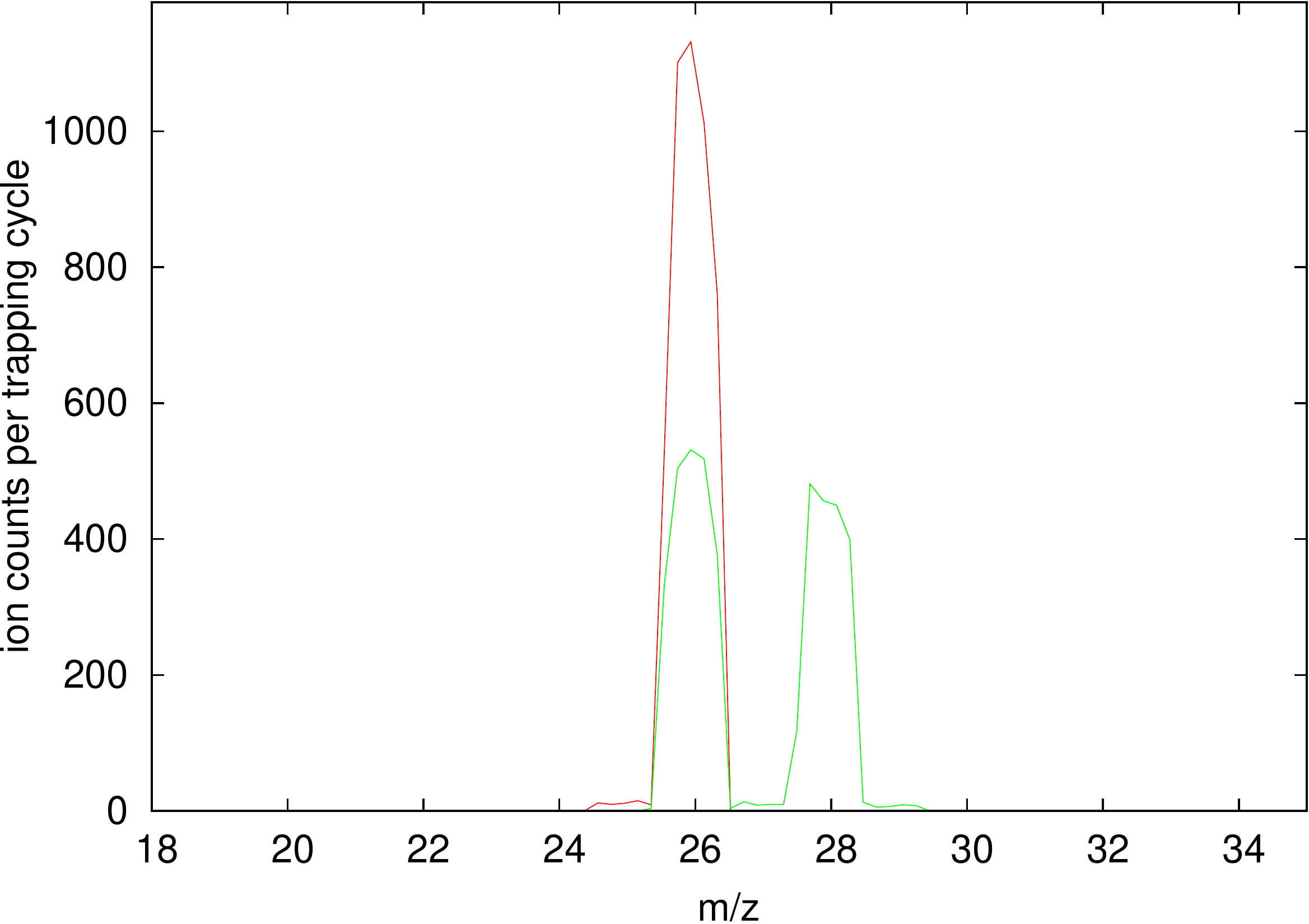}
 \caption{\label{fig3} Mass spectra taken 700~ms after injection of mass selected CN$^+$ ions
 into the 15~K cold ion trap; the red and green traces are recorded with and without 
 admission of H$_2$ ($n \approx 8 \times 10^{10}$~cm$^{-3}$), respectively.
 For the latter case, the ions undergo on average 86 collisions with hydrogen molecules.
While CN$^+$ reacts swiftly to \ce{HCNH+} ($m/z=28$), 
about half of the injected ions with $m/z=26$ show a very slow reaction 
with H$_2$, pointing to a contamination on mass 26~u (see text).}
\end{figure}
Fig.~\ref{fig3} shows an example measurement with and without 
the admission of a low concentration of H$_2$ to the trap.
While we could confirm that at low temperature the reaction chain is 
indeed proceeding close to the collision rates given above  
(we refrain from giving exact values for $k_2$ and $k_3$ because a precise pressure 
calibration is not implemented in the trapping machine yet), 
it also became evident that about half of the mass-selected ion ensemble admitted to the trap 
is reacting only very slowly with H$_2$. This finding points to the fact that the $m/z=26$ ions
admitted to the trap do not comprise CN$^+$ exclusively but
are most likely heavily contaminated with another isobaric species
(such a conclusion is not possible based on the LIICG spectra shown in 
Fig.~\ref{fig2}).  
Based on the ion mass and the \ce{CH3CN} precursor, 
we suspect the acetylene cation, HCCH$^+$, 
to be the ion in question (or the H$_2$C$_2^+$ radical cation quickly isomerizing to HCCH$^+$).
The acetylene cation is known to react very slowly with H$_2$  at the low temperatures
prevailing in our ion trap~\cite{ger93,sor94}.
Although CH$_3$CN can be handled very safely
and turned out to be an ideal precursor for the spectroscopic experiments presented in this work, 
in future kinetic and spectroscopic experiments, it might be advantageous
to use HCN (heavily diluted in He) as precursor gas in the ion source to 
produce cleaner samples of CN$^+$ and HCN$^+$/HNC$^+$~\cite{pet90}.
In particular, we plan accurate low-temperature measurements of $k_2$ and $k_3$ with 
the approach outlined in Refs.~\cite{asv04,asv04a}.
Also,  HCN$^+$/HNC$^+$ will be a prime spectroscopic target
in coming investigations, 
as no high-resolution  data whatsoever exist to date.




\section*{Acknowledgments}
This work (including the research visit of JLD in K{\"o}ln) 
has been  supported by the Deutsche
Forschungsgemeinschaft (DFG) via SFB 956 project B2  (project ID 18401886)
and DFG SCHL 341/15-1 (“Cologne Center for Terahertz Spectroscopy”).
JLD acknowledges partial support from the Spanish AEI through grant
FIS2016-77726-C3-1-P and from the European Research Council through grant
agreement ERC-2013-SyG-610256-NANOCOSMOS. JLD expresses his immense 
gratitude to F. J. Lovas, and to current and past 
members of the former “Molecular Physics Division” at NBS, then 
“Molecular Physics Division” at NIST, now “Optical Technology 
Division” at NIST, for their continuous inspiration, mentor-ship 
and friendship.

\section*{Appendix A. Supplementary material}
Supplementary data associated with this article can be found, in 
the online version, at XXX.



\bibliographystyle{elsarticle-num}         
\bibliography{LIRTRAP,sthorwirth_bibdesk,extraref,CN+}           

\end{document}